# Space, time and altruism in pandemics and the climate emergency

Chris T. Bauch[1,*], Athira Satheesh Kumar[1,2], Kamal Jnawali[3], Karoline Wiesner[4], Simon A. Levin[5], and Madhur Anand[2]

[1]Department of Applied Mathematics, University of Waterloo, Canada
[2]School of Environmental Sciences, University of Guelph, Canada
[3]Department of Mathematics, State University of New York at Oswego, USA
[4]Institute of Physics and Astronomy, University of Potsdam, Germany
[5]Department of Ecology and Evolutionary Biology, Princeton University, USA
[*]cbauch@uwaterloo.ca

## Abstract

Climate change is a global emergency, as was the COVID-19 pandemic. Why was our collective response to COVID-19 so much stronger than our response to the climate emergency, to date? We hypothesize that the answer has to do with the scale of the systems, and not just spatial and temporal scales but also the 'altruistic scale' that measures whether an action must rely upon altruistic motives for it to be adopted. We treat COVID-19 and climate change as common pool resource problems that exemplify coupled human-environment systems. We introduce a framework that captures regimes of containment, mitigation, and failure to control. As parameters governing these three scales are varied, it is possible to shift from a COVID-like system to a climate-like system. The framework replicates both inaction in the case of climate change mitigation, as well as the faster response that we exhibited to COVID-19. Our cross-system comparison also suggests actionable ways that cooperation can be improved in large-scale common pool resources problems, like climate change. More broadly, we argue that considering scale and incorporating human-natural system feedbacks are not just interesting special cases within non-cooperative game theory, but rather should be the starting point for the study of altruism and human cooperation.

# 1 Introduction

## 1.1 Background

The Intergovernmental Panel for Climate Change (IPCC) called for rapid, unprecedented social change to address the climate emergency [1, 2]. However, progress has been inadequate, despite the risk to lives and economies. In contrast, the COVID-19 pandemic–also a global emergency–stimulated rapid, unprecedented social change in a very short period of time [3, 4, 5], even to the extent of reducing global greenhouse gas emissions by 14% during shutdowns [6]. This difference in humanity's collective response calls for an explanation [7].

To address this question, we treat climate change and infectious diseases as public goods problems, where everyone in a population benefits equally from a good that must be maintained by voluntary contributions from individual members. Specifically, both a stable climate and a disease-free state are 'nonexcludable' (everyone benefits equally from a stable climate or a disease-free state) and 'nonrivalrous' (one person benefiting from a stable climate or the absence of disease does not prevent someone else from also benefiting) [8, 9]. Hence, a stable climate and a disease-free state are public goods.

In the case of climate change, some mitigation measures require individuals to adopt behavior with a real or perceived cost, such as installing photovoltaic panels on their roof, or walking instead of driving a car [1, 10]. However, climate change mitigation benefits everyone (not just the mitigator) and hence there is a temptation to 'free-ride' by enjoying a stable climate without adopting costly mitigation measures [11, 12, 13]. Similarly, herd immunity against an infectious disease benefits everyone, but the reluctance of some individuals to follow public health measures that involve a real or perceived cost (such as mask use or correct use of antibiotics) can lead to resurgence of the infection, thus harming everyone [14, 9, 15, 16]. Even though mitigating climate change or an epidemic is best for the group as a whole, non-cooperating might be beneficial at the individual level, so that free-riding harms the public good [17, 18]. Other examples of public goods problems include air quality, water quality, and biodiversity [19]. Everyone benefits from clean air or water, but individuals and industries may pollute without bearing the full cost of the negative impacts [20].

However, other factors can help prevent free-riding and change the game from a Prisoner's Dilemma to a Coordination game, where coordination can allow all players to choose mitigation. This can occur if a penalty is applied to non-mitigation, such as through injunctive social norms [21] or other forms of punishment [22], if decisions occur across multiple hierarchical levels [23] or can unfold through time instead of being instantaneous [22], or if the players think that a dangerous tipping point might occur [24], among other possible mechanisms.

## 1.2 The role of scales

While both climate change and infectious disease control are public goods problems, they are dissimilar in many respects, such as spatial scale. The greenhouse effect involves a well-mixed system (the atmosphere) where emissions from a single tailpipe are dispersed planet-wide after a relatively short period of time, whereas diseases transmitted from person to person spread locally. An epidemic begins when infection control in the initial stages are a failure, due either to properties of the pathogen, lack of knowledge, or lack of adherence to control measures (the latter of which is our particular focus). If the contacts of Patient Zero act quickly to contain an outbreak, then the epidemic is prevented and the larger population is spared costly mitigation measures. Action by a few individuals can thus prevent exponentially higher morbidity and mortality, and the requirement to adopt costly measures, for those further down the transmission chain. However, if some contacts fail to act and become infected, they may in turn pass the infection to their own set of contacts. As a result, the infection percolates through the entire population and eventually jumps to other populations. This effect has been termed the 'race to trace' in the context of ring vaccination for smallpox [25]. In contrast to the smallpox endgame, we lost the 'race to trace' during COVID-19 due to behavioral and pathogen features, causing us to resort to mass mitigation.

Climate change and infectious diseases also have very different temporal scales. Epidemics caused by acute respiratory viruses pass through a population in a matter of weeks [26, 27], while the impacts of greenhouse gas emissions are manifested over generations [28, 29, 30, 31]. Contending with the human tendency to discount outcomes far in the future is a continuing challenge for climate change mitigation.

Finally, there is a difference in what could be called the 'altruistic scale' of climate change mitigation and infectious disease control. If a single individual chooses to adopt a climate mitigation measure, the direct benefit to the individual of the resulting reduction in GHG emissions is extremely small and close to zero, thus the 'personal efficacy'–a measure of direct individual benefit–is low. (This neglects other social or psychological factors that might support mitigation behavior, of course [32].) However, an individual adopting a pandemic mitigation measure receives a relatively larger direct benefit through protection from infection, since the efficacy of masks and vaccines against infection are significantly higher. Because the personal efficacy of mitigation measures are higher for infectious diseases, their adoption is less reliant on altruistic motivations [33, 34]. Essentially, for complex coupled human-environment systems, we suggest that the (clinical) efficacy of an intervention should be framed in the context of whether the intervention's uptake relies upon altruistic motives, versus simply relying upon directly personal benefit to the mitigator.

In the same vein, it is worth pointing out that in both systems, the benefit of the mitigator's behavior might be different for the mitigator and the others in the population [35], according to environmental, health and economic vulnerability. Individuals in low-income countries are more impacted by climate change, and thus more benefited by climate change mitigation. For an infectious disease intervention to protect others in the population, it faces a double barrier: it has to be effective both in protecting the mitigator, and also protective in terms of the mitigator not passing on the infection to their contacts. Some vaccines prevent disease in the vaccinated individual but do not block transmission. Whereas, masking protects the masker and also prevents onward transmission. However, a masker who gets infected might reduce their morbidity while also passing on the virus to a more vulnerable contact [36].

Our objective is to develop a network-based framework to help us understand how individual and group behavior in public goods problems depend on the spatial, temporal, and altruistic scales of the system. Exploring these three scales provides a structured way to examine and compare behavior in these systems and thus better understand the difference in population reactions to COVID-19 versus the climate crisis. Comparing pandemic and climate systems can help address why mitigating the pandemic was partly successful even if the race to trace was lost, whereas mitigating climate change remains a challenge [1, 37, 38, 2, 39]. Although both climate change and pandemics are comprised of multiple and very different systems, by 'Climate Change' we will mean a rise in the mean global temperature anomaly through anthropogenic release of greenhouse gases, and we will use 'COVID-19' to refer to a fully susceptible population that has been seeded by an acute respiratory virus transmitted from person to person, such as for COVID-19 or influenza, for instance.

## 2 Results

### 2.1 Theoretical Framework

We hypothesize that variation across three basic dimensions explains many differences in population responses to COVID-19 versus the climate crisis (Figure 1).

- Spatial scale: How widely is a change in individual behavior felt in the natural system? How widely do changes in the natural system impact individuals?

- Temporal scale: How long does it take for individual behavior change to be felt in the natural system? How quickly do changes in the natural system impact individuals? In other words, how long do populations wait to experience the feedback of their collective decisions?

- Altruistic scale: How much direct benefit accrues to the individual choosing to adopt mitigation measures? Is altruistic motivation required for mitigation to work? How are the benefits of their action distributed across the population?

We illustrate the role of these dimensions through a network consisting of nodes and links. We consider a network of $N$ individuals (nodes), where each individual has a certain number of neighbors (links, $Q$). Individuals choose whether or not to adopt mitigation behavior. This behavior impacts the abundance of a natural agent that spreads through network links (e.g., viral shedding in the case of infectious diseases, or greenhouse gas concentration in the case of climate change). The natural agent decays at some rate per unit time ($\delta$). Individuals decide whether to adopt mitigation measures based on a utility function that is determined, in part, by the personal efficacy ($\varepsilon$), representing how effective the behavior is in directly protecting the individual. Other parameters also play a role: the rate of spread of the agent through links is governed by transmission parameter ($R$). Adopting control measures imposes a cost on the mitigator ($c_M$), and a person impacted by infection or climate change also suffers a cost ($c_I$). There is an average duration of time that individuals spend in the impacted state ($D$). Nodes that are exposed to the agent suffer impact to their well-being with a specified probability ($\beta$). A complete description of the simulation model is in the Supplementary Appendix. We use the terms 'COVID-19' and 'Climate change' to describe model scenarios, but these are meant to represent idealized systems, as opposed to being highly realistic descriptors of severe acute respiratory syndrome coronavirus 2 (SARS-CoV-2) dynamics, or the global temperature anomaly caused by greenhouse gas emissions.

Each of the three scales–space, time and altruism–is governed by a model parameter: neighborhood size $Q$ (for space), decay rate $\delta$ (for time) and personal efficacy $\varepsilon$ (for altruism). These parameters differ by orders of magnitude between the two systems. Greenhouse gas emissions caused by a single person spread globally, so the relevant spatial scale is a fully connected network where every person is connected to every other person ($Q = N - 1$). In contrast, diseases that are transmitted from person to person spread to a relatively small number of neighbors ($Q \ll N$). For climate change, the decay rate $\delta$ is low, since greenhouse gases persist for a long time in the atmosphere, but for acute respiratory infections that resolve in days or weeks, $\delta$ is high. Finally, the incremental benefit to a single individual mitigator due to reducing GHG emissions is very small because the resulting incremental change in worldwide GHG emissions is tiny, hence $\varepsilon \approx \frac{1}{N} \approx 0$. Infectious disease interventions can provide a significant benefit to the individual mitigator who adopts them, regardless of whether they are effective in protecting others or how large the population size is, hence $0 \ll \varepsilon < 1$. An agent-based network simulation describing this framework appears in the Supplementary Appendix.

## 2.2 Containment, mitigation and non-control

This framework captures regimes of strong mitigation, and failure to mitigate (non-control), for both study systems. For COVID-19, with a relatively low neighborhood size $Q$, fast decay rate $\delta$, and high personal efficacy $\varepsilon$, strong mitigation is easier to obtain, and it occurs even when the cost of mitigation ($c_M$) is relatively high (Figure 2, top). However, in the case of climate change, achieving strong mitigation is more challenging. The population is completely connected ($Q \approx N - 1$) and hence spatial localization does not occur–the population is homogeneously mixing. Personal efficacy $\varepsilon$ is likewise very small, so the number of individuals who choose not to mitigate grows rapidly. Climate change can only be prevented if the cost of mitigation $c_M$ is very small (Figure 2, bottom). (Or, should the cost of mitigation become negative, such as when renewable sources of energy cost less than carbon-based sources, then strong mitigation is achievable.)

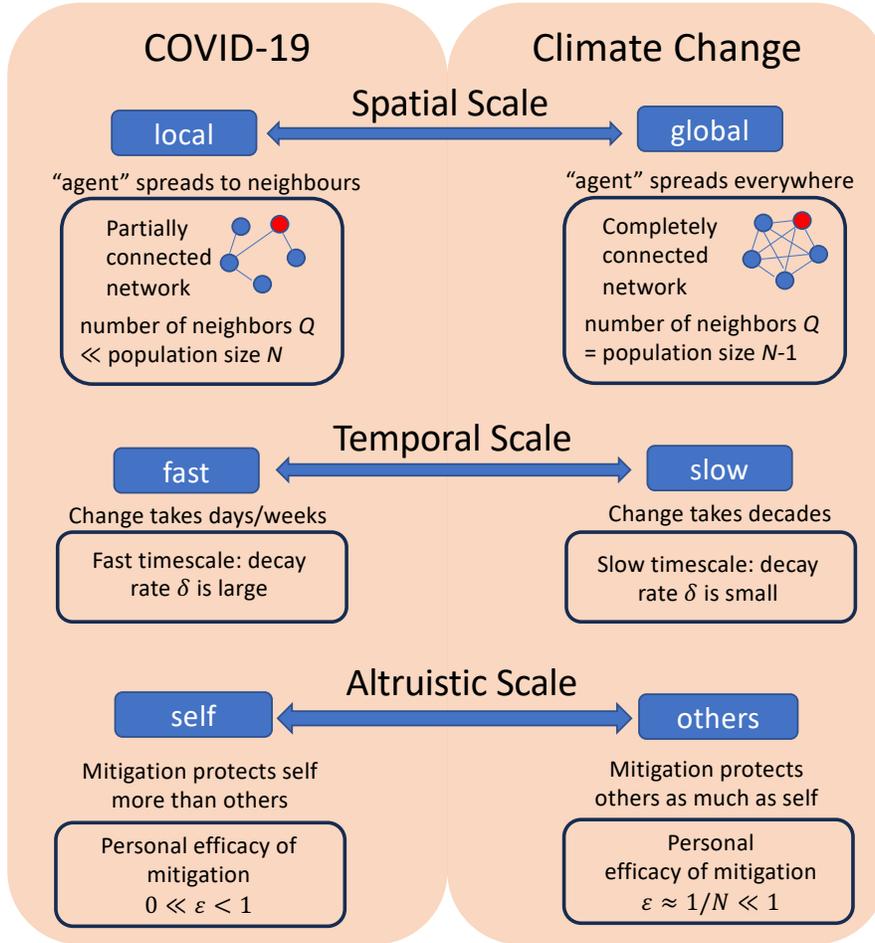

Figure 1: Schematic representation of spatial, temporal, and altruistic scales in COVID-19 and climate change. The two systems differ significantly with respect to each of these scales.

Closer study of the framework suggests why these differences occur, and also reveals a third dynamical regime of containment. Consider a hypothetical scenario where a small number of non-mitigators are distributed throughout the population, and they are emitting the natural agent at some rate. How do their neighbors react? In the containment regime, a sufficient proportion of neighbors immediately choose mitigation, thus completely blocking the epidemic/onset of climate change. Whereas, in the mitigation regime, many neighbors will choose to mitigate, but some others will not, thus allowing the spread of the natural agent into new areas of the network and causing percolation of both agent and partial mitigation throughout the network. Alternatively, many neighbors may choose to mitigate, but the personal efficacy is just not high enough to prevent continued spread of the agent. Finally, in the non-control regime, almost no neighbors choose mitigation, and the natural agent spreads widely, impacting all individuals in the network.

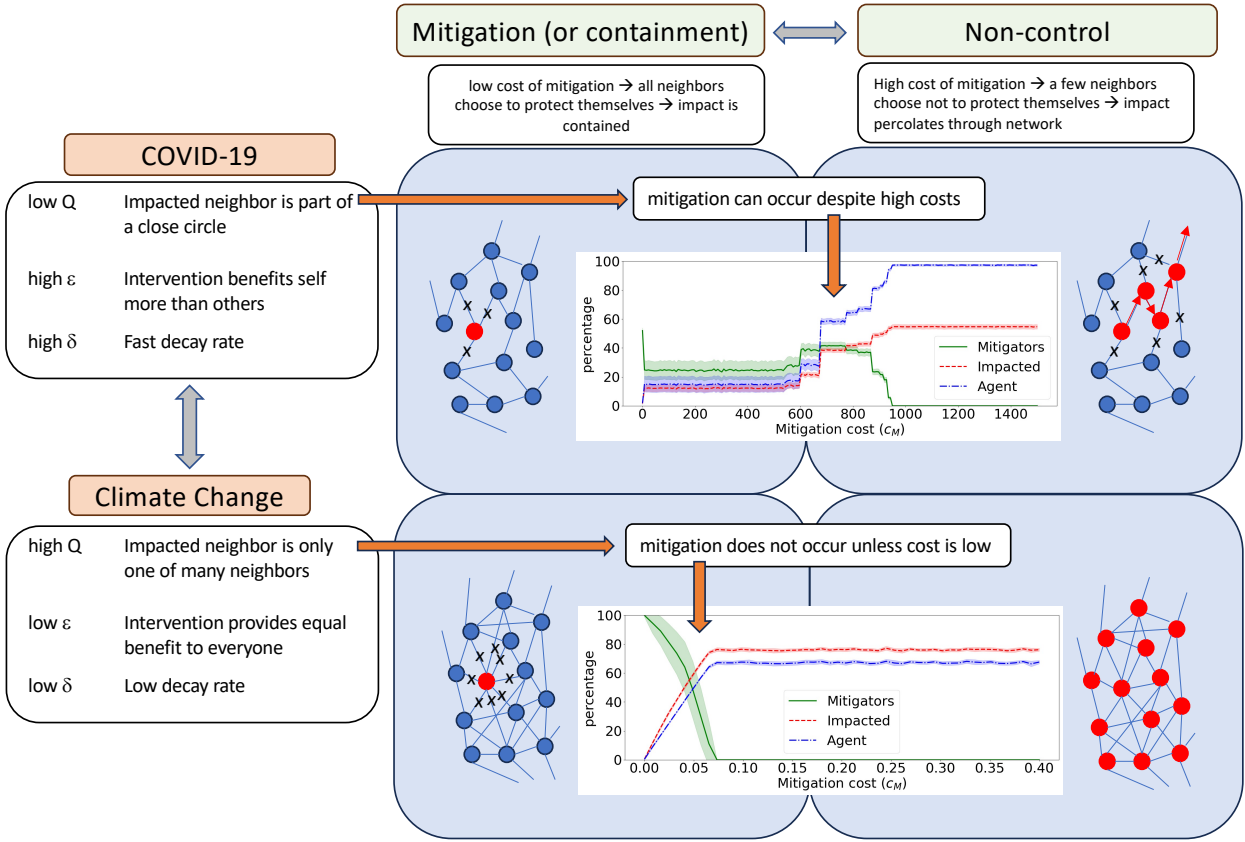

Figure 2: Containment, mitigation, and non-control regimes for COVID-19 and climate change. The cost of mitigation has to be significantly lower for the climate change system, in order to facilitate the uptake of mitigating behavior. Parameters are in the Supplementary Appendix: Table 1. Shaded regions denote one standard deviation of simulation results at their steady state.

This intuition is borne out in simulation results of the agent-based network model. In the case of COVID-19, when mitigation measures are not too costly, the neighbors of any initial non-mitigator nodes are more likely to adopt mitigation, since that single non-mitigator neighbor is a clear and present threat representing a considerable share of a node's total number of contacts, and the personal efficacy is high enough to motivate a choice to mitigate. Thus, all neighbors of the non-mitigator adopt mitigation, and the agent is completely contained (Figure 3 top left). Simulations in this regime show that the population responds quickly to surges in the agent prevalence, acting to contain the agent and causing it to decline once again. The result is cycles of epidemic outbreaks [40] and behavioral responses [41], similar to those seen in many infectious disease systems (Figure 3 top left). This regime covers a broad portion of the $c_M$ parameter space, on account of the strong negative feedback loop that containment represents. However, if the mitigation cost is somewhat larger, a few neighbors will choose not to mitigate. This allows percolation of the agent through the network, and a suboptimal outcome of both higher levels of agent and higher levels of mitigation (Figure 3 top middle). The population loses the 'race to trace' [25], and the paradoxical (and suboptimal) outcome is that both mitigation behavior and agent prevalence are more widespread. The transition between these two dynamical regimes is sudden and shows tipping point behavior, on account of the fact that the negative feedback loop of containment can be overcome by the positive feedback loop of agent propagation, once the cost of mitigation becomes too high. In fact, the model simulation shows several such regime shifts as $c_M$ is increased, reflecting the additional role played by the small-world network geometry (Figure 3 middle). If the mitigation cost is very large, no neighbors will choose to mitigate, and the non-control regime is obtained, with few mitigators and a high number of impacted nodes (Figure 3, top right).

For climate change, a containment regime does not occur at our baseline parameter values, although a mitigation regime is possible at a sufficiently low cost of mitigation (Figure 3 middle). There is a very small regime of cycles in behavior, although the temporal evolution of the natural agent is too slow for the cycles to be very pronounced in the agent prevalence time series (Figure 3 bottom left). For higher values of the mitigation cost, mitigation fails altogether and the agent spreads without hindrance (Figure 3 bottom right). In both COVID-19 and climate change, the success of mitigation depends on the time scale of mitigating behavior being at least as fast as the timescale of agent spread. When agent spreads as fast as mitigating behavior evolves, a tight feedback loop is possible. If the spread of agent is much slower than the uptake of mitigation, then behavior always adapts quickly enough to keep the agent under control. But slow behavioral change allows the agent to spread faster than the population can adapt its behavior.

It is possible to move the system between a COVID-like regime of containment/strong mitigation and a climate-like regime of weak mitigation, by varying the three scaling parameters for the neighborhood size $Q$, decay rate $\delta$, and personal efficacy $\varepsilon$ (Figure 4). As the neighborhood size increases, the population structure changes from a localized network where containment occurs, to a globally connected population in the non-control regime (Figure 4a). As the neighborhood size increases, the neighbors of non-mitigators are less incentivized to adopt mitigation on account of the diminishing direct impact to them individually. When the personal efficacy ($\varepsilon$) is very high, containment is possible on account of the intervention being attractive to individuals, and effective as a means to prevent agent spread (Figure 4b). However, for lower values of $\varepsilon$ representing an imperfect intervention, the agent is able to percolate through imperfect rings of mitigator neighbors. As a result, both mitigation behavior and agent prevalence are higher [42]. However, when $\varepsilon$ is very small, mitigation becomes too unattractive to justify the cost, and the population abandons it altogether. Finally, at high values of the decay rate, it is relatively easy to contain the agent, but at lower values of the decay rate, the agent persists for long periods of time and this makes it difficult to contain, necessitating more widespread mitigation (Figure 4c). Transitions in the timescale parameter $\delta$ result in a more gradual switch between the systems. In the case of climate change, these regimes correspond to the expected initial failure to take up climate change mitigation on account of low $\varepsilon$, followed by partial adoption of mitigation as the impacts grow over time (Figure 4b, $\varepsilon \approx 0.05$). If the decay rate $\delta$ is small, eventually there are enough greenhouse gases to force more of the population to choose mitigation, although by this time the agent and its impacts are already widespread (Figure 4c, $\delta \approx 0.01$).

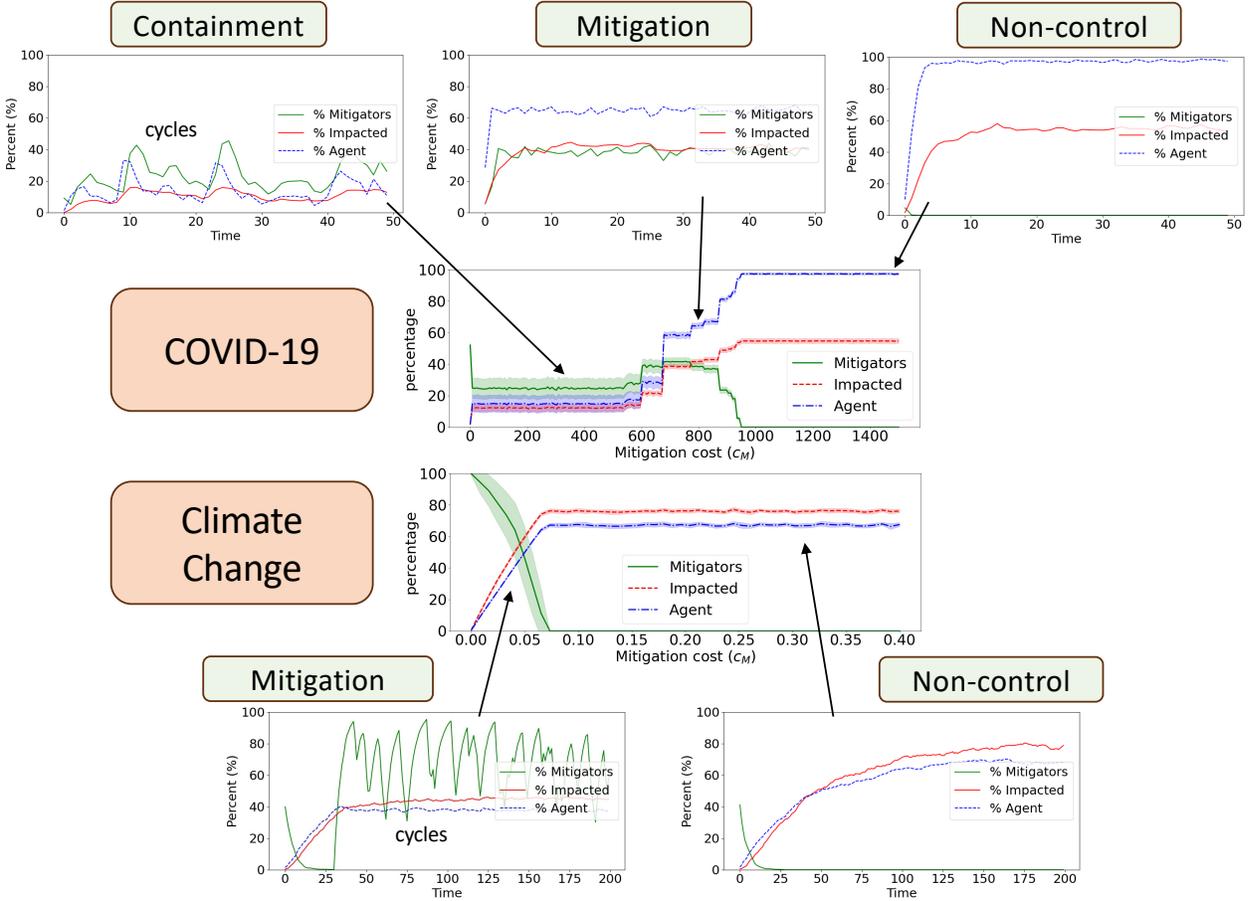

Figure 3: Temporal dynamics. COVID-19 exhibits a broad regime of containment, and cycles in both behavior and agent prevalence. Climate change only allows for a narrow regime of mitigation where pronounced cycles are possible for behavior but not agent prevalence. Parameters are in the Supplementary Appendix: Table 1. Shaded regions denote one standard deviation of simulation results at their steady state.

## 3 Discussion

The study of why humans cooperate and how cooperation can be sustained is a question of enduring scholarly interest [43, 44]. We think that both (1) system scale and (2) natural system feedbacks are indispensable to the study of human cooperation. With respect to the role of scale, it has been noted previously how humans continue to cooperate in large populations even in one-shot interactions [45, 44], and despite the fact that our cooperative tendencies evolved in small hunter-gatherer groups [46]. Whether human cooperation can be sustained in a globalized society–and for how long–is an existential question that can be answered by explicitly addressing the role of scale. With respect to the role of resource dynamics and natural system feedbacks, we note that most of the theory of cooperation has been develop for fixed payoff matrices where the decisions of players do not impact the payoff they receive [47]. This is a good first approximation. However, most of the things that humans compete for in the real world are of limited availability and can be depleted, which is why they are a source of competition in the first place. Thus, we argue that studying cooperation in coupled human-environment systems [48, 49], and especially the role of scale in these systems [50, 51], is not merely a special case of non-cooperative game theory, but rather should be its starting point and primary framework.

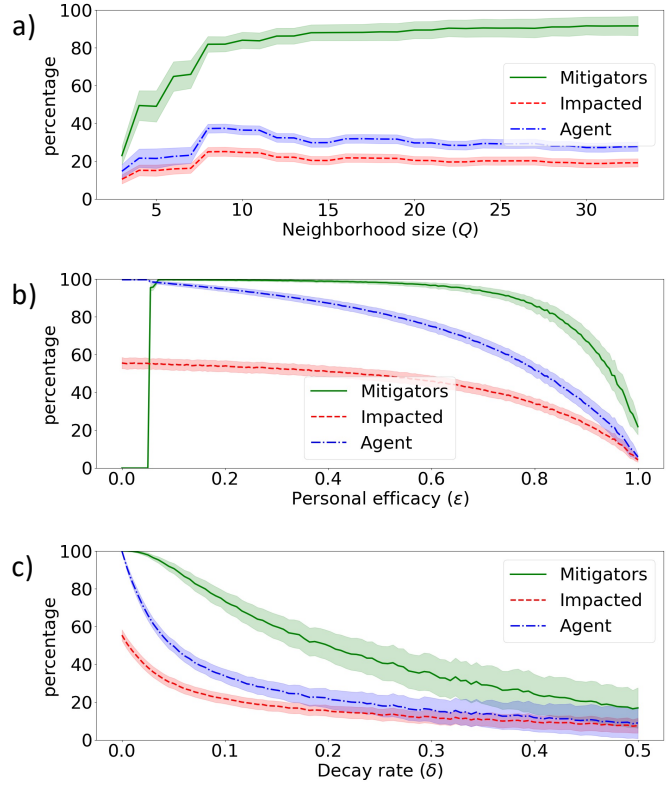

Figure 4: Effect of varying the three primary scaling parameters on dynamical regime, for (a) neighborhood size $Q$, (b) decay rate $\delta$, and (c) personal efficacy $\varepsilon$. As the parameter are varied, regime changes from mitigation to non-control. The step structure of subpanel (a) is due to geometry of the network for small values of $Q$. Other parameters are in the Supplementary Appendix: Table 1. Shaded regions denote one standard deviation of simulation results at their steady state.

The effects of spatial and temporal scales on systems is likewise a frequent topic of study [52]. However, the impact of these scales on public goods problems is less studied [53, 54, 46] and comparative analysis for specific system types is rare. Here we proposed an explanation for the widely differing reaction to the emergencies of COVID-19 and climate change, in terms of not only spatial and temporal scales, but also an altruism scale that quantifies the relative extent to which a behavior benefits self or others. The concept of an altruism scale emerges naturally once common pool resource (CPR) problems are moved into a human-environment framework, where we are forced to consider how actions affect a natural system, which in turn affects the human population again, in a two-way feedback loop [55, 48, 49]. As an example of practical applications, we argue that epidemiological studies could estimate not only clinical (personal) efficacy of an intervention, but also estimate the 'social efficacy'–how well an intervention protects others, especially for transmissible conditions like infectious diseases or obesity [56, 35].

Our framework identifies the cost of mitigation as one of the main determinants of mitigating behavior, for both climate change and COVID-19. However, the population may adopt control measures at a higher cost for COVID-19 than for climate change. This reflects how individuals during the first wave of COVID-19 were willing to adopt not only contact precautions like handwashing, mask wearing and social distancing, but also more costly measures such as school and workplace closures. In contrast, control or even mitigation against climate change requires a very low cost to control or mitigate, on account of the large neighborhood size, slow decay rate, and low personal efficacy. If this low cost can be achieved, then the switch to mitigation behavior is predicted to be quick. This switch might be accelerated by a change from net costs for renewable energy to net savings relative to carbon-based energy sources, for example [57].

Our simulation model necessarily makes simplifying assumptions that could affect its predictions. For instance, we assume that individuals have perfect knowledge of their impact risk, and that decisions are based solely on parameters such as utilities, mitigation costs and cost of impact. We also intentionally modelled the natural system dynamics in similar ways for COVID-19 and climate change, aside from differences in the parameter values used. Although the scale of time, space and altruism are important, other factors can be crucial in CPRs, such as the level of organization complexity of the system [58]. For instance, better resource management happens under combined community and governmental efforts [59, 60]. In addition to sustainability, CPR management focuses on multiple factors, such as government and community interactions along with cultural, economic, and ecological factors, by looking at the interaction between human and environmental systems [61, 62]. Similarly, there is a strong influence on human decisions by the choice of their neighbors, especially in the case of environmental decision-making [63, 64]. The framework could be extended in this direction, which would also benefit from the extensive literature on complex network structure [65, 66] and especially the role of social processes on networks [67, 68]. Our primary objective was to demonstrate the value of comparing CPR problems with different characteristic scales, as opposed to creating a realistic model of networks for either system.

Finally, future work could formally quantify the 'altruism scale'. Here, we simply equated personal efficacy with clinical efficacy and changed the interpretation of the parameter, but future work could explicitly distinguish between personal efficacy (degree of self protection) versus social efficacy (degree of protection to others), how individual decision-making is affected by both, and how they differ for interventions in various coupled human-environment systems. This could allow using the framework to understand when and how other-regarding preferences can be a motivator for mitigation, such as when considering the multi-generational impacts of climate change [54, 69], or the desire to protect immediate family and vulnerable members of the community through contact precautions, during the COVID-19 [70]. The role of spatial locality in promoting altruism and cooperation (through providing benefits to likely kin) is already recognized in the literature on 'viscosity' [71] and here we are suggesting that the modes of operation of specific intervention types (vaccines, masking, emission reductions) could benefit from closer attention in this body of literature. In a similar vein, the NIMBY effect (Not In My Back Yard) exemplifies a conflict between global benefits and local costs similar (though not identical) to that of climate change, that could be studied with our framework.

This framework also suggests possible strategies to accelerate the uptake of climate change mitigation. For instance, the framework suggests that framing a problem as local instead of global (smaller $Q$) could help support uptake. For climate change, this could occur through emphasizing co-benefits of reduced GHG emissions, such as air pollution reduction [72]. Approaching climate change as a local problem can also leverage faster behavioral change by considering the local and shorter-term impacts (high $\delta$) of extreme weather events like heatwaves, floods, and droughts. In conclusion, comparing common pool resources problems with very different spatial, temporal, and altruistic scales may not only explain differences in human behavior but also suggests ways in which more sustainable behaviors could be encouraged.

# References

[1] VP Masson-Delmotte, Panmao Zhai, SL Pirani, C Connors, S Péan, N Berger, Y Caud, L Chen, MI Goldfarb, and Pedro M Scheel Monteiro. Ipcc, 2021: Summary for policymakers. in: Climate change 2021: The physical science basis. contribution of working group i to the sixth assessment report of the intergovernmental panel on climate change. 2021.

[2] Tejal Kanitkar, Akhil Mythri, and T Jayaraman. Equity assessment of global mitigation pathways in the ipcc sixth assessment report. *Climate Policy*, pages 1–20, 2024.

[3] Marco Ciotti, Massimo Ciccozzi, Alessandro Terrinoni, Wen-Can Jiang, Cheng-Bin Wang, and Sergio Bernardini. The covid-19 pandemic. *Critical reviews in clinical laboratory sciences*, 57(6):365–388, 2020.

[4] Teck Ling, Gabriel Hoh, Chyong Ho, and Christina Mee. Effects of the coronavirus (covid-19) pandemic on social behaviours: From a social dilemma perspective. *Technium Soc. Sci. J.*, 7:312, 2020.

[5] Jason Shachat, Matthew J Walker, and Lijia Wei. How the onset of the covid-19 pandemic impacted pro-social behaviour and individual preferences: Experimental evidence from china. *Journal of Economic Behavior & Organization*, 190:480–494, 2021.

[6] Vineet Singh Sikarwar, Annika Reichert, Michal Jeremias, and Vasilije Manovic. Covid-19 pandemic and global carbon dioxide emissions: A first assessment. *Science of the Total Environment*, 794:148770, 2021.

[7] Lorraine Whitmarsh, Wouter Poortinga, and Stuart Capstick. Behaviour change to address climate change. *Current Opinion in Psychology*, 42:76–81, 2021.

[8] Alessandro Tavoni, Astrid Dannenberg, Giorgos Kallis, and Andreas Löschel. Inequality, communication, and the avoidance of disastrous climate change in a public goods game. *Proceedings of the National Academy of Sciences*, 108(29):11825–11829, 2011.

[9] Caroline E Wagner, Joseph A Prentice, Chadi M Saad-Roy, Luojun Yang, Bryan T Grenfell, Simon A Levin, and Ramanan Laxminarayan. Economic and behavioral influencers of vaccination and antimicrobial use. *Frontiers in Public Health*, 8:614113, 2020.

[10] A Damodaran. *India, Climate Change, and the Global Commons*. Oxford University Press, 2023.

[11] David Andersson, Sigrid Bratsberg, Andrew K Ringsmuth, and Astrid S de Wijn. Dynamics of collective action to conserve a large common-pool resource. *Scientific Reports*, 11(1):9208, 2021.

[12] Joshua J. Lawler, Timothy H. Tear, Christopher R. Pyke, M. Rebecca Shaw, Patrick Gonzalez, Peter Kareiva, Lara J. Hansen, Lee Hannah, Kirk R. Klausmeyer, Allison Aldous, Craig Bienz, and Sam H. Pearsall. 1. resource management in a changing and uncertain climate. *Frontiers in Ecology and the Environment*, 2010.

[13] Gherardo Girardi and Gian Lorenzo Preite. Escaping the economist's straightjacket: Overcoming the free-rider mentality which prevents climate change from being effectively addressed. *Climate Change Research at Universities: Addressing the Mitigation and Adaptation Challenges*, pages 561–575, 2017.

[14] Marc Lipsitch and Matthew H Samore. Antimicrobial use and antimicrobial resistance: a population perspective. *Emerging infectious diseases*, 8(4):347, 2002.

[15] Katinka den Nijs, Jose Edivaldo, Bas DL Châtel, Jeroen F Uleman, Marcel Olde Rikkert, Heiman Wertheim, and Rick Quax. A global sharing mechanism of resources: modeling a crucial step in the fight against pandemics. *International Journal of Environmental Research and Public Health*, 19(10):5930, 2022.

[16] Luojun Yang, Sara M Constantino, Bryan T Grenfell, Elke U Weber, Simon A Levin, and Vítor V Vasconcelos. Sociocultural determinants of global mask-wearing behavior. *Proceedings of the National Academy of Sciences*, 119(41):e2213525119, 2022.

[17] Douglas D Heckathorn. The dynamics and dilemmas of collective action. *American sociological review*, pages 250–277, 1996.

[18] Chris T Bauch and David JD Earn. Vaccination and the theory of games. *Proceedings of the National Academy of Sciences*, 101(36):13391–13394, 2004.

[19] Joyeeta Gupta, Xuemei Bai, Diana M Liverman, Johan Rockström, Dahe Qin, Ben Stewart-Koster, Juan C Rocha, Lisa Jacobson, Jesse F Abrams, Lauren S Andersen, et al. A just world on a safe planet: a lancet planetary health–earth commission report on earth-system boundaries, translations, and transformations. *The Lancet Planetary Health*, 8(10):e813–e873, 2024.

[20] William Rees. What's blocking sustainability? human nature, cognition, and denial. *Sustainability: Science, Practice and Policy*, 6(2):13–25, 2010.

[21] Tamer Oraby, Vivek Thampi, and Chris T Bauch. The influence of social norms on the dynamics of vaccinating behaviour for paediatric infectious diseases. *Proceedings of the Royal Society B: Biological Sciences*, 281(1780):20133172, 2014.

[22] Jobst Heitzig, Kai Lessmann, and Yong Zou. Self-enforcing strategies to deter free-riding in the climate change mitigation game and other repeated public good games. *Proceedings of the National Academy of Sciences*, 108(38):15739–15744, 2011.

[23] Vadim A Karatayev, Vítor V Vasconcelos, Anne-Sophie Lafuite, Simon A Levin, Chris T Bauch, and Madhur Anand. A well-timed shift from local to global agreements accelerates climate change mitigation. *Nature communications*, 12(1):2908, 2021.

[24] Scott Barrett and Astrid Dannenberg. Climate negotiations under scientific uncertainty. *Proceedings of the National Academy of Sciences*, 109(43):17372–17376, 2012.

[25] Edward H Kaplan, David L Craft, and Lawrence M Wein. Emergency response to a smallpox attack: the case for mass vaccination. *Proceedings of the National Academy of Sciences*, 99(16):10935–10940, 2002.

[26] Xinyin Xu, Jing Zeng, Runyou Liu, Yang Liu, Xiaobo Zhou, Lijun Zhou, Ting Dong, Yuxin Cha, Zhuo Wang, Ying Deng, et al. Should we remain hopeful? the key 8 weeks: spatiotemporal epidemic characteristics of covid-19 in sichuan province and its comparative analysis with other provinces in china and global epidemic trends. *BMC infectious diseases*, 20:1–15, 2020.

[27] Lori Post, Kasen Culler, Charles B Moss, Robert L Murphy, Chad J Achenbach, Michael G Ison, Danielle Resnick, Lauren Nadya Singh, Janine White, Michael J Boctor, et al. Surveillance of the second wave of covid-19 in europe: longitudinal trend analyses. *JMIR public health and surveillance*, 7(4):e25695, 2021.

[28] Jonathan White. Climate change and the generational timescape. *The Sociological Review*, 65(4):763–778, 2017.

[29] Kira Vinke, Sabine Gabrysch, Emanuela Paoletti, Johan Rockström, and Hans Joachim Schellnhuber. Corona and the climate: a comparison of two emergencies. *Global Sustainability*, 3:e25, 2020.

[30] JS Maloy. Beyond crisis and emergency: Climate change as a political epic. *Ethics & International Affairs*, 38(1):103–125, 2024.

[31] Tim Bickerstaffe. A problem of generations? habitus, social processes and climate change. *Journal of Global Responsibility*, 15(1):111–124, 2024.

[32] Cameron Brick, Anna Bosshard, and Lorraine Whitmarsh. Motivation and climate change: A review. *Current Opinion in Psychology*, 42:82–88, 2021.

[33] Meng Li, Eric G Taylor, Katherine E Atkins, Gretchen B Chapman, and Alison P Galvani. Stimulating influenza vaccination via prosocial motives. *PloS one*, 11(7):e0159780, 2016.

[34] Cornelia Betsch, Robert Böhm, and Lars Korn. Inviting free-riders or appealing to prosocial behavior? game-theoretical reflections on communicating herd immunity in vaccine advocacy. *Health Psychology*, 32(9):978, 2013.

[35] Arne Traulsen, Simon A Levin, and Chadi M Saad-Roy. Individual costs and societal benefits of interventions during the covid-19 pandemic. *Proceedings of the National Academy of Sciences*, 120(24):e2303546120, 2023.

[36] Derek K Chu, Elie A Akl, Stephanie Duda, Karla Solo, Sally Yaacoub, Holger J Schünemann, Amena El-Harakeh, Antonio Bognanni, Tamara Lotfi, Mark Loeb, et al. Physical distancing, face masks, and eye protection to prevent person-to-person transmission of sars-cov-2 and covid-19: a systematic review and meta-analysis. *The lancet*, 395(10242):1973–1987, 2020.

[37] Carmela Gulluscio, Pina Puntillo, Valerio Luciani, and Donald Huisingh. Climate change accounting and reporting: A systematic literature review. *Sustainability*, 12(13):5455, 2020.

[38] Hoesung Lee, Katherine Calvin, Dipak Dasgupta, Gerhard Krinner, Aditi Mukherji, Peter Thorne, Christopher Trisos, José Romero, Paulina Aldunce, and Alexander C

Ruane. Climate change 2023 synthesis report summary for policymakers. *CLIMATE CHANGE 2023 Synthesis Report: Summary for Policymakers*, 2024.

[39] Anna Pirani, Jan S Fuglestvedt, Edward Byers, Brian O'Neill, Keywan Riahi, June-Yi Lee, Jochem Marotzke, Steven K Rose, Roberto Schaeffer, and Claudia Tebaldi. Scenarios in ipcc assessments: lessons from ar6 and opportunities for ar7. *npj Climate Action*, 3(1):1, 2024.

[40] David JD Earn, Pejman Rohani, Benjamin M Bolker, and Bryan T Grenfell. A simple model for complex dynamical transitions in epidemics. *science*, 287(5453):667–670, 2000.

[41] Chris T Bauch. Imitation dynamics predict vaccinating behaviour. *Proceedings of the Royal Society B: Biological Sciences*, 272(1573):1669–1675, 2005.

[42] Xingru Chen and Feng Fu. Imperfect vaccine and hysteresis. *Proceedings of the royal society B*, 286(1894):20182406, 2019.

[43] Michael Tomasello. *Why we cooperate*. MIT press, 2009.

[44] Samuel Bowles and Herbert Gintis. Origins of human cooperation. *Genetic and cultural evolution of cooperation*, 2003:429–43, 2003.

[45] David G Rand and Martin A Nowak. Human cooperation. *Trends in cognitive sciences*, 17(8):413–425, 2013.

[46] Robert Boyd and Peter J Richerson. Punishment allows the evolution of cooperation (or anything else) in sizable groups. *Ethology and sociobiology*, 13(3):171–195, 1992.

[47] Michael Doebeli and Christoph Hauert. Models of cooperation based on the prisoner's dilemma and the snowdrift game. *Ecology letters*, 8(7):748–766, 2005.

[48] Clinton Innes, Madhur Anand, and Chris T Bauch. The impact of human-environment interactions on the stability of forest-grassland mosaic ecosystems. *Scientific reports*, 3(1):2689, 2013.

[49] Billie Lee Turner, Pamela A Matson, James J McCarthy, Robert W Corell, Lindsey Christensen, Noelle Eckley, Grete K Hovelsrud-Broda, Jeanne X Kasperson, Roger E Kasperson, Amy Luers, et al. Illustrating the coupled human–environment system for vulnerability analysis: three case studies. *Proceedings of the National Academy of Sciences*, 100(14):8080–8085, 2003.

[50] Karoline Wiesner, Jyotsna Puri, and Andreas Reumann. Â€ complexity-awareâ€™ monitoring and evaluation of development programs â€" anchoring them in complexity science. *Advances in Complex Systems (ACS)*, 27(06):1–17, 2024.

[51] J. Ladyman and K. Wiesner. *What Is a Complex System?* Yale University Press, 2020.

[52] Benjamin M Bolker and Stephen W Pacala. Spatial moment equations for plant competition: understanding spatial strategies and the advantages of short dispersal. *The American Naturalist*, 153(6):575–602, 1999.

[53] Kurt Erik Schnier. Spatial externalities and the common-pool resource mechanism. *Journal of Economic Behavior & Organization*, 70(1-2):402–415, 2009.

[54] Henrik Österblom and Øyvind Paasche. Earth altruism. *One Earth*, 4(10):1386–1397, 2021.

[55] Alison P Galvani, Chris T Bauch, Madhur Anand, Burton H Singer, and Simon A Levin. Human–environment interactions in population and ecosystem health. *Proceedings of the National Academy of Sciences*, 113(51):14502–14506, 2016.

[56] Nicholas A Christakis and James H Fowler. The spread of obesity in a large social network over 32 years. *New England journal of medicine*, 357(4):370–379, 2007.

[57] Rupert Way, Matthew C Ives, Penny Mealy, and J Doyne Farmer. Empirically grounded technology forecasts and the energy transition. *Joule*, 6(9):2057–2082, 2022.

[58] Simon A Levin. The problem of pattern and scale in ecology: the robert h. macarthur award lecture. *Ecology*, 73(6):1943–1967, 1992.

[59] Zhiqi Zhang, Xiangyu Jia, Zeren Gongbu, Dingling He, and Wenjun Li. Common pool resource governance in strong-government context: A case study of caterpillar fungus (ophiocordyceps sinensis) on the qinghai-tibet plateau. *Environmental Science & Policy*, 152:103644, 2024.

[60] Leticia Antunes Nogueira, Karin Andrea Wigger, and Suyash Jolly. Common-pool resources and governance in sustainability transitions. *Environmental innovation and societal transitions*, 41:35–38, 2021.

[61] Bhim Adhikari. The economics of common pool resources: A review. *Ecology, Economy and Society–the INSEE Journal*, 4(1):71–88, 2021.

[62] Chengyi Tu, Renfei Chen, Ying Fan, and Xuwei Pan. Impact of resource availability and conformity effect on sustainability of common-pool resources. *arXiv.org*, abs/2310.07577, 2023.

[63] Annika M Wyss, Sebastian Berger, and Daria Knoch. Pro-environmental behavior in a common-resource dilemma: The role of beliefs. *Journal of Environmental Psychology*, 92:102160, 2023.

[64] Samuel Bowles and Herbert Gintis. A cooperative species: Human reciprocity and its evolution. In *A Cooperative Species*. Princeton University Press, 2011.

[65] Andrea Avena-Koenigsberger, Joaquín Goñi, Ricard Solé, and Olaf Sporns. Network morphospace. *Journal of the Royal Society Interface*, 12(103):20140881, 2015.

[66] Jonathan F Donges, Yong Zou, Norbert Marwan, and Jürgen Kurths. Complex networks in climate dynamics: Comparing linear and nonlinear network construction methods. *The European Physical Journal Special Topics*, 174(1):157–179, 2009.

[67] Damon Centola, Robb Willer, and Michael Macy. The emperor's dilemma: A computational model of self-enforcing norms. *American Journal of Sociology*, 110(4):1009–1040, 2005.

[68] Kevin JS Zollman. Social network structure and the achievement of consensus. *Politics, Philosophy & Economics*, 11(1):26–44, 2012.

[69] José Manuel Ortega-Egea, Nieves García-de Frutos, and Raquel Antolín-López. Why do some people do "more" to mitigate climate change than others? exploring heterogeneity in psycho-social associations. *PLoS One*, 9(9):e106645, 2014.

[70] Carl L Hanson, Ali Crandall, Michael D Barnes, and M Lelinneth Novilla. Protection motivation during covid-19: A cross-sectional study of family health, media, and economic influences. *Health Education & Behavior*, 48(4):434–445, 2021.

[71] Sebastien Lion and Minus van Baalen. Self-structuring in spatial evolutionary ecology. *Ecology letters*, 11(3):277–295, 2008.

[72] Matt S Sparks, Isaiah Farahbakhsh, Madhur Anand, Chris T Bauch, Kathryn C Conlon, James D East, Tianyuan Li, Megan Lickley, Fernando Garcia-Menendez, Erwan Monier, et al. Health and equity implications of individual adaptation to air pollution in a changing climate. *Proceedings of the National Academy of Sciences*, 121(29):e2215685121, 2024.

# Supplementary Appendix: Space, time and altruism in pandemics and the climate emergency

Chris T. Bauch[1,*], Athira Satheesh Kumar[1,2], Kamal Jnawali[3], Karoline Wiesner[4], Simon A. Levin[5], and Madhur Anand[2]

[1]Department of Applied Mathematics, University of Waterloo, Canada
[2]School of Environmental Sciences, University of Guelph, Canada
[3]Department of Mathematics, State University of New York, USA
[4]Institute of Physics and Astronomy, University of Potsdam, Germany
[5]Department of Ecology and Evolutionary Biology, Princeton University, USA
[*]cbauch@uwaterloo.ca

## 1 Supplementary Material

### 1.1 Network model description

To instantiate our framework, we built an agent-based simulation of a small-world network [1] of $N$ individuals, each with an average of $Q$ neighbours. Each node has a state corresponding to their behavior, well-being, and presence or absence of agent:

- The behavioral state of node $i$ is either mitigator ($M$) or non-mitigator ($N$) and is denoted $B_i \in \{M, N\}$. A node can change behavior in each timestep depending on how attractive mitigation or non-mitigation appears to them at that time.

- The state of well-being of node $i$ is either well ($W$) or impacted ($I$) and is denoted $H_i \in \{W, I\}$. Impacted individuals have been infected, or have suffered a climate change impact, but they recover after some period of time.

- The agent status indicating whether agent is present (1) or absent (0) at node $i$ is denoted $A_i \in \{1, 0\}$. Agent can be transferred to or from neighbouring nodes, and also decays. The presence of agent at a node can cause impacts.

Individuals can move from impacted back to well according to a parameter $D$ governing the duration of impact. The prevalence of the natural agent that node $i$ is subject to in a given timestep is denoted $A_i^{risk}$ (which is the local prevalence of virus for an infectious disease, or greenhouse gas (GHG) concentration for climate change). This quantity influences the probability of impact. Individuals transition between behavioral states with a specified probability per timestep, as described in the following subsections.

## 1.2 Behavioral transitions

The utility for node $i$ by choosing mitigating behavior is given by

$$e_{i,N} = -p_i^N c_I D \tag{3}$$

whereas the utility for not mitigating is

$$e_{i,M} = -p_i^M c_I D - c_M \tag{4}$$

where $c_M$ is the cost of being a mitigator, $c_I$ is the cost of being impacted, $D$ is the duration of the impact, $p_i^M$ is the probability of being impacted if node $i$ is a mitigator, and $p_i^N$ is the probability of being impacted if node $i$ is a non-mitigator. The probabilities $p_i^M$ and $p_i^N$ are given by:

$$p_i^M = \beta(1-\varepsilon)(1 - e^{-RA_i^{risk}}) \tag{5}$$

$$p_i^N = \beta(1 - e^{-RA_i^{risk}}) \tag{6}$$

where $0 < \beta < 1$ controls the probability if impact (analogous to morbidity risk), $0 < \varepsilon < 1$ is the personal efficacy of mitigation, $R$ controls how rapidly the agent is transmitted through the network, and $A_i^{risk} = \sum_j A_j / \sum_j$ is the sum over the neighbours of $i$ and represents the proportion of neighbours of node $i$ that carry agent (agent status '1'). This represents the local risk to agent $i$. Note how $\varepsilon$ reduces the risk of being impacted due to local prevalence $A_i^{risk}$, but mitigators must also pay a cost $c_M$ for that protection.

In each timestep, with a probability $0 < \sigma < 1$, a node decides whether to changes its behavior. A node that is currently not mitigating will switch to mitigation if

$$e_{i,M} > e_{i,N}$$

while a node that is mitigating will switch to non-mitigation if

$$e_{i,N} > e_{i,M}$$

## 1.3 Physical transitions

If agent is present at a node, it decays with probability $\delta$ per timestep, changing the node's agent status from '1' to '0'. In each timestep, a non-mitigator that currently has agent status '0' changes to agent status '1' with a generation probability $p_{gen}$, indicating agent is now present. This corresponds to individual behavior that is responsible for greenhouse gas emissions, or the spread of infection during the acute infection phase when the virus is multiplying in the host. The agent can also be transmitted to other nodes, depending on whether participating nodes are mitigators, who attempt to protect themselves from spread according to the personal efficacy $\varepsilon$. In particular, if an agent's current status is '0', it will receive agent from its neighbours with probabilities

$$p_i^{inf,M} = (1-\varepsilon)(1 - e^{-RA_i^{risk}}) \tag{5}$$

if they are a mitigator, and

$$p_i^{inf,N} = (1 - e^{-RA_i^{risk}}) \tag{6}$$

if they are a non-mitigator. We note that agent status is imperfectly analogous to the infection state in compartmental epidemic models [2], whereas the impacted state corresponds to a state of disease/morbidity caused by an infection.

## 1.4 Well-being transitions

A currently 'well' node with agent (new or existing) becomes impacted with probability $\beta$, and an impacted node recovers with probability $1/D$ per timestep, where $D$ is the average duration in the impacted state.

## 1.5 Simulation design

In each timestep, the following updates occur:

1. Generation, decay, and recovery: running through each node in the population from $i = 0$ to $i = N$,

    - A non-mitigating node with agent status '0' transitions to agent status '1' with probability $p_{gen}$.
    - A node with agent status '1' transitions to '0' with probability $\delta$.
    - A node with an impacted status becomes well with probability $1/D$.

2. Behaviour: running through each node in the population from $i = 0$ to $i = N$,

    - Determine local risk to each node, $A_i^{risk}$.
    - With probability $\sigma$, the node decides whether to change its behaviour. A non-mitigator switches to mitigator if $e_{N,i} > e_{M,i}$, and a mitigator swiches to non-mitigator if $e_{N,i} < e_{M,i}$. (No change if they are exactly equal.)

3. Agent spread, and impacts: running through each node in the population from $i = 0$ to $i = N$,

- A node with agent status '0' transitions to agent status '1' due to spread from neighbours, with probabilities $p_i^{inf,M}$ or $p_i^{inf,N}$ depending on their behavioral state.
- A node with agent status '1' transitions from healthy to impacted with probability $\beta$.
- All agent states are updated simultaneously according to these transition events.

We initialize a population where 1% of the nodes carry the agent, 50% of the population are mitigators, and 100% of the population are healthy. For COVID-19, our baseline values assumed a small neighborhood ($Q$), a rapid decay of the natural agent ($\delta$) and a high personal efficacy ($\varepsilon$). Similarly, other timescale variables $R$, $D$ and $p_\sigma$ were relatively large for COVID-19, reflecting the faster timescales of infectious disease spread. For climate change, our baseline values assumed a neighborhood size equal to the entire population ($Q = N - 1$), a slow decay rate ($\delta$), and a low personal efficacy ($\varepsilon$), while $R$, $D$ and $p_\sigma$ were relatively small. Other parameter values were the same for the baseline climate and COVID-19 scenarios. Baseline parameter values appear in Table 1). The simulation was coded in Python (see Supplementary File).

| Parameters | Description | Values |
|---|---|---|
| $Q$ | Neighbourhood size (node degree) (COVID/Climate) | 5 / 999 |
| $\varepsilon$ | Personal efficacy of mitigation (COVID/Climate) | 0.95 / 0.01 |
| $\delta$ | Agent decay probability, per timestep (COVID/Climate) | 0.5 / 0.01 |
| $R$ | Proportionality constant for rate of spread (COVID/Climate) | 5 / 0.001 |
| $D$ | Duration of impact, in timesteps (COVID/Climate) | 5 / 50 |
| $p_\sigma$ | Probability of changing behavior (COVID/Climate) | 0.9 / 0.2 |
| $c_M$ | Cost of being a mitigator | 50 |
| $c_I$ | Cost of being impacted | 1000 |
| $\beta$ | Impact probability | 0.2 |
| $p_{gen}$ | Agent generation probability, per timestep | 0.02 |
| $N$ | Total number of individuals | 1000 |
| Function | Description | |
| $A_i^{risk} = \sum_j A_j / \sum_j$ | proportion of neighbours of node $i$ that carry agent | |
| $p_i^{inf,M} = (1-\varepsilon)e^{-RA_i^{risk}}$ | Probability of mitigating node $i$ receiving agent from neighbors | |
| $p_i^{inf,N} = e^{-RA_i^{risk}}$ | Probability of non-mitigating node $i$ receiving agent from neighbors | |
| $p_i^M = \beta p_i^{inf,M}$ | Probability of node $i$ being impacted, if a mitigator | |
| $p_i^N = \beta p_i^{inf,N}$ | Probability of node $i$ being impacted, if a non-mitigator | |
| $e_{i,M} = c_M - p_i^M c_I D$ | Payoff to node $i$ for being a mitigator | |
| $e_{i,N} = -p_i^N c_I D$ | Payoff to node $i$ for being a non-mitigator | |

Table 1: Baseline model parameters and functions. All parameters are dimensionless, and the model is discrete in time.

# References

[1] Duncan J Watts and Steven H Strogatz. Collective dynamics of 'small-world' networks. *nature*, 393(6684):440–442, 1998.

[2] Herbert W Hethcote. The mathematics of infectious diseases. *SIAM review*, 42(4):599–653, 2000.